\documentclass[12pt]{article}
\usepackage{amsfonts}
\usepackage{stmaryrd}
\usepackage{graphicx}
\usepackage{mathrsfs}
\usepackage{picinpar}
\usepackage[all]{xy}
\usepackage{pifont}
\usepackage{latexsym}   
\usepackage{amsmath,  amssymb,  amsthm,  latexsym}
\usepackage{txfonts}
\usepackage{cite}
\usepackage[colorlinks,linkcolor=red,anchorcolor=blue,citecolor=blue]{hyperref}
\begin{document}
\newtheorem{lemma}{Lemma}[section]
\newtheorem{proposition}{Proposition}[section]
\newtheorem{theorem}{Theorem}[section]
\newtheorem{corollary}{Corollary}[section]
\newtheorem{example}{Example}[section]
\newtheorem{definition}{Definition}[section]
\newtheorem{remark}{Remark}[section]
\newtheorem{property}{Property}[section]
\newtheorem{postulate}{Postulate}[section]

\title{\Large\bf  Exogenous Quantum Operator Logic Based on Density Operators}
\author{Yunguo Lin$^{ \mbox{{\small a, b}}}$
\qquad {Yongming Li$^{ \mbox{{\small a, }}}$\thanks{Corresponding author. E-mail address: liyongm@snnu.edu.cn(Y.Li)}}
\qquad \\\\$^{\mbox{{\small a}}}${\small College of Computer Science, Shaanxi Normal University, Xi'an, 710062, China}
\\ $^{\mbox{{\small b}}}${\small College of Computer and Information Sciences,}
\\ {\small Fujian Agriculture and Forestry University, Fuzhou, Fujian 350002, China}}

\date{}
\maketitle
\begin{center}
\begin{minipage}{140mm}
\centerline{\bf Abstract} \vskip 3mm {\qquad }

\ \ \ \ \ Although quantum logic by using exogenous approach has been
proposed for reasoning about closed quantum systems, an improvement
would be worth to study quantum logic based on density operators instead
of unit vectors in the state logic point of view. In order to achieve this,
we build an exogenous quantum operator logic({\bf EQOL})based on density operators for reasoning about open quantum systems. We show that this logic is sound and complete. Just as the exogenous quantum propositional logic({\bf EQPL}), by applying exogenous approach, {\bf EQOL} is extended from the classical propositional logic, and is used to describe the state logic based on density operators. As its applications, we confirm the entanglement property about Bell states by reasoning and logical argument, also verify the existence of eavesdropping about the basic BB84 protocol. As a novel type of mathematical formalism for open quantum systems, we introduce an exogenous quantum Markov chain({\bf EQMC}) where its quantum states are labelled using {\bf EQOL} formulae. Then, an example is given to illustrate the termination verification problem of a generalized quantum loop program described using {\bf EQMC}.

\end{minipage}
\end{center}

\vskip 4mm

\section{Introduction}

Since the 1990s, with the implementation of Shor algorithm for factorizing the large integers and the Grover quantum search algorithm, quantum theory is widely accepted as a successful theory of nature science\cite{[1]}. As an aspect of quantum theory's development, in a series of recent papers, a good deal of work about quantum logic has been discussed\cite{[2],[3],[4],[5],[6]}. There are at least two kinds of quantum logics for quantum systems. One is called quantum logics developed by Birkhoff and von Neumann\cite{[7],[8],[9],[10]}. Another one is called quantum computation logics for quantum information systems, which are developed for quantum computation and quantum information\cite{[7],[11],[12]}. The difference between quantum logics and logics for quantum information system is: the formers concern about a basic semantic question with an emphasis on high abstract level mathematical structures, i.e., the lattices of closed subspaces of a Hilbert space where the classical connectives by new connectives representing the lattices operations, while the latters specially focus on their applications and are described in the framework of quantum computation and quantum information.

Unlike the mainstream quantum logic, a novel logic is introduced in\cite{[13],[14],[15],[16],[17],[18]} for modeling and reasoning about quantum systems. It is an extension of probability logic by using exogenous approach, looks like modal logic which is an extension of the classical proposition logic. The logic is very powerful and specialized for describing quantum mechanical components and procedures. It can reason about a finite collection of qubits, and express quantum states in a Hilbert space. So, it is suitable to apply in quantum computation and information.

The exogenous approach doesn't change the models of the original logic, and only adds some additional structures on collections of those models as they are\cite{[18]}. For more details, we refer to R.Chadaha\cite{[18]}. By using the exogenous approach,  R.Chadaha et al. adopt models of classical proposition logic as their original models, give quantum models(semantics) with superpositions of classical models(valuations), and design a logic language for constraining these superpositions. They call it an exogenous quantum propositional logic(\textbf{EQPL})\cite{[18]}, and prove that \textbf{EQPL} is sound and complete. Besides, for describing and reasoning about the evolution of quantum system, they also propose several dynamic logics and temporal logics as extensions of exogenous quantum propositional logic. In particular, they introduce quantum linear time logic and quantum computational tree logic, provide their weak completeness proofs, and study their satisfiability and the model checking problems\cite{[14],[17]}.

Now, logics for quantum systems by using exogenous approach have been widely applied in the model checking problems about quantum protocols and communicating quantum processes\cite{[19],[20],[21],[22],[23],[24]}. P.Mateus et al. have investigated the model checking problems for exogenous temporal quantum logics, and reasoned about the BB84 protocol by using their logics\cite{[14]}. Tim Davidson et al. have introduced a quantum model checker which is used to verify the correction and safety of quantum protocols\cite{[24]}. They specify properties of quantum protocols using exogenous quantum computation tree logic.

As indicated in \cite{[17],[18]}, there are many other problems to be done along exogenous approach. In the exogenous quantum proposition logic, the state logic is based on the unit vectors, and so \textbf{EQPL} is used to reason about the closed quantum systems. An improvement is to build a quantum logic based on the density operators as the state logic which is suitable to describe open quantum systems. We have noted that a term language in \textbf{EQPL} consists of amplitude terms, probability terms, alternation terms, etc., and is interpreted in the real closed field of the quantum structure or in the closure of the real closed fields. These elements are used to describe the probability characteristics occurring in measurements. But, this term language doesn't seem to describe the density operators. Hence, by changing the term language, it may be possible to achieve the desired logic. Meanwhile, we also hope that the desired logic can be used to describe quantum communication, quantum cryptographic protocols, quantum programs in open quantum environment. To these ends, in this paper, we will build an quantum operator logic in an exogenous perspective. It is our idea and the major technical contribution of this paper. To be specific, we propose the following.
\begin{enumerate}
  \item[(1)] In accordance with the fundamental idea of exogenous approach, we still keep classical propositional language unchanged, including its semantics. It is taken as an original fragment of our logic.
  \item[(2)] In the view of the quantum measurement, we define a operator term language which consists of some kinds of operator terms. These operator terms are interpreted in a collection of super-operators on $H_{qB}$. We show that the role of these operator terms is similar to those of \textbf{EQPL}.
  \item[(3)] We denote a comparison between two operator terms as a quantum atomic proposition, then recursively define exogenous quantum operator formulae as global formulae, and eventually build exogenous quantum operator logic(\textbf{EQOL}). The quantum atomic proposition is a key notion which indirectly gives a comparison between the probabilities of two outcomes occurring in measurement.
  \item[(4)] We introduce a quantum operator structure based on a projective measurable space which enables us to propose the semantic of our logic. As a logical system, we show our logic is sound and complete.
  \item[(5)] As its applications, we research the entanglement property about Bell states, and the the existence of eavesdropping about the basic BB84 protocol. Also, we introduce a novel notion of quantum Markov chain which is used to illustrate the termination verification problem of a generalized quantum loop program.
\end{enumerate}

The rest of the paper is organised as follows. We recall the four postulates of quantum mechanics based on density operators, and introduce several quantum operators, a projective measurable space in Section 2. We present the syntax, semantic, quantum operator formulae of \textbf{EQOL} in Section 3. We show the soundness of \textbf{EQOL} in Section 4 and the completeness of \textbf{EQOL} in Section 5. We illustrate \textbf{EQOL} with several examples about the Bell states, the BB84 protocol, a novel notion of quantum Markov chain and a generalized quantum Loop programs in Section 6. We summarize our results and the future work in Section 7.

\section{Notations and Preliminaries}
For the convenience of the reader, we write some basic notions that are needed in this paper.

\subsection{Basic notations}
We only consider finite dimensional Hilbert spaces. Let $qB$ be a finite set of qubit symbols, we write $qB=\{qb_1,qb_2,\cdots,qb_n\}$. For $A\in 2^{qB}$, $A$ is a valuation on $qB$ where $qb$ is true if $qb\in A$, otherwise, it is false. $2^{qB}$ is a collection of all valuations on $qB$. The Hilbert space $H_{qB}$ is spanned by $2^{qB}$ which constitutes a standard computational basis, i.e., $H_{qB}=span\{|v\rangle|v\in 2^{qB}\}$. $S(H_{qB})$ is a collection of super-operators on $H_{qB}$. $P(H_{qB})=\{P\in S(H_{qB}):P=P^{*}=P^2\}$ is a collection of projective measurement operators, and $P_v\in P(H_{qB})$ denotes a projective measurement operator on the subspace spanned by $v\in 2^{qB}$. $D(H_{qB})$ is a collection of all density operators.

\subsection{Four postulates}
In the exogenous quantum propositional logic\cite{[18]}, the unit vectors are taken to guide the state logic for closed quantum systems. But, to deal with the open and composed quantum systems, we need to consider the density operators. In this section, we recall the four postulates of quantum mechanics in density operator picture\cite{[1]} which will guide our design logic, and also briefly introduce several basic ideas and concepts. For more details, we refer to Nielsen and Chuang\cite{[1]}.

\begin{postulate}
Associate to any isolated physical system is a complex vector space with a Hilbert space known as the state space of the system. The system is completely described by its density operator which is a positive operator $\rho$ with trace one, acting on the state space of the system. If a quantum system is in the state $\rho_i$ with probability $p_i$, then the density operator for the system is $\Sigma_ip_i\rho_i$.
\end{postulate}

In  R.Chadaha's work, a qubit state is a superposition of two valuations $|0\rangle$ and $|1\rangle$ of a classical bit. Furthermore, a $n-$qubits state is a superposition of those classical valuations of $n-$classical bits. Accordingly, in density operator picture, a mixed state is a probability distribution(ensemble) $\{p_i,|\psi_i\rangle\}$ on pure states(classical valuations) $\{|\psi_i\rangle\}$ with probability $0<p_i\leq 1$, $\Sigma_{i=1}^np_i=1$.

\begin{postulate}
The evolution of a closed quantum system is described by a unitary
transformation. That is, the state $\rho$ of the system at time $t_1$ is related to the state $\rho'$ of the system at time $t_2$ by a unitary operator $U$ which depends only on the times $t_1$ and $t_2$, that is, $\rho'=U\rho U^{\dagger}$.
\end{postulate}

\begin{postulate}
Quantum measurements are described by a collection $\{P_v\}$ of measurement operators. These are operators acting on the state space of the system being measured. The index $v$ refers to the measurement outcomes that may occur in the experiment. If the state of the quantum system is $\rho$ immediately before the measurement then the probability that result $v$ occurs is given by $p(v)=tr(P_v\rho P^{\dagger}_v)$, and the state of the system after measurement is $\frac{P_v\rho P^{\dagger}_v}{tr(P_v\rho P^{\dagger}_v)}$. The measurement operators satisfy the completeness equation, i.e., $\Sigma_vP^{\dagger}_vP_v=I$.
\end{postulate}

Let $V$ be all possible outcomes, the Postulate 2.3 denotes that a possible outcome $v\in V$ is observable at the mixed state $\rho$ with $tr(P_v\rho P^{\dagger}_v)$. We assume that $V$ is a collection of all classical valuations of $qB$($n-$qubits), that is $V=2^{qB}$, then a density operator $\rho$ is a probability distribution $\{tr(P_v\rho P^{\dagger}_v), |v\rangle\}$ of all classical valuations by using measurement operators $\{P_v\}$. In fact, this result is similar to a probability distribution of all possible outcomes occurring in quantum measurements at a quantum pure state.

\begin{postulate}
The state space of a composite physical system is a tensor product of the state spaces of the component physical systems. Moreover, if we have systems numbered 1 through $n$, and system number $i$ is prepared in the state $\rho_i$, then the joint state of the total system is $\rho_1\otimes\rho_2\otimes\cdots\otimes\rho_n$.
\end{postulate}

Without loss of generality, we assume that a physical system $H_{qB}$ is a tensor product of two component physical systems $H^1_{qB}$ and $H^2_{qB}$. Let $\rho_1\in D(H^1_{qB})$, $\rho_2\in D(H^2_{qB})$, then $\rho=\rho_1\otimes\rho_2\in D(H_{qB})$. In accordance with the postulate 2.3, every density operator $\rho_i$ is a probability distribution $\{tr(P_{v^i}\rho P^{\dagger}_{v^i}),|v^i\rangle\}$, $i=1,2$, then the density operator $\rho$ can be expressed as a probability distribution $\{tr(P_w\rho P^{\dagger}_w),|w\rangle\}$, where $w$ is a string concatenation of $v^1$ and $v^2$, $|w\rangle$ is a tensor product of $|v^1\rangle$ and $|v^2\rangle$.

\subsection{Projective measurable space}

The aim of this paper is to propose an exogenous quantum logic based on density operators for open quantum systems, in which the term of our logic is described not by the probability terms, the amplitude terms, etc., but by the operators which are interpreted in the super-operators $S(H_{qB})$. Using these operators as quantum measurements, we will indirectly explain the probability characteristics of outcomes occurring in measurements at a density operator. To this end, we need a concept be similar with the probabilistic measurable space. We will extend the probabilistic measurable space to a projective measurable space in which the probabilities are replaced by the super-operators.

Let $S(H_{qB})$ be the set of super-operators on $H_{qB}$. Both $(S(H_{qB}),0,+)$ and $(S(H_{qB}),$ $I,\cdot)$ are monoids, where $0$ and $I$ are the null and identity super-operators on $H_{qB}$. Let $\varepsilon_1,\varepsilon_2\in S(H_{qB})$, for any $\rho\in D(H_{qB})$, $(\varepsilon_1\cdot \varepsilon_2)(\rho)=\varepsilon_1(\varepsilon_2(\rho))$, $(\varepsilon_1+\varepsilon_2)(\rho)=\varepsilon_1(\rho)+\varepsilon_2(\rho)$. We omit the symbol $\cdot$ and write $\varepsilon_1\varepsilon_2$ for $\varepsilon_1\cdot \varepsilon_2$. We can show that $(S(H_{qB}),+,\cdot)$ forms a semring\cite{[25]}.

\begin{definition}
Let $\varepsilon_1,\varepsilon_2\in S(H_{qB})$,

\emph{(1)} $\varepsilon_1\lesssim_{\rho}\varepsilon_2$ if for a given $\rho\in D(H_{qB})$, $tr(\varepsilon_1(\rho))\leq tr(\varepsilon_2(\rho))$, also write by $\varepsilon_2\gtrsim_{\rho}\varepsilon_1$ or $\varepsilon_2\lnsim_{\rho} \varepsilon_1$;

\emph{(2)} $\varepsilon_1\lesssim \varepsilon_2$ if for any $\rho\in D(H_{qB})$, $tr(\varepsilon_1(\rho))\leq tr(\varepsilon_2(\rho))$;

\emph{(3)} $\varepsilon_1\eqsim_{\rho}\varepsilon_2$ if for a given $\rho\in D(H_{qB})$, $tr(\varepsilon_1(\rho))=tr(\varepsilon_2(\rho))$;

\emph{(4)} $\varepsilon_1\eqsim \varepsilon_2$ if for any $\rho\in D(H_{qB})$, $tr(\varepsilon_1(\rho))=tr(\varepsilon_2(\rho))$.
\end{definition}

In this definition, the trace $tr$ at a quantum state $\rho$ is the probability that the quantum state is reached. $\varepsilon_1\lesssim_{\rho}\varepsilon_2$($\varepsilon_1\lesssim \varepsilon_2$)is used to compare the ability of trace preservation. We denote them that the probability of measurement outcomes occurring in the projective measurement $\varepsilon_1$ is always not greater than that of performing $\varepsilon_2$, for a given $\rho\in D(H_{qB})$(any $\rho\in D(H_{qB})$).

\begin{definition}
Let $(V,\mathscr{P}V)$ be a measurable space, that is, $V$ is a set of classical valuations and $\mathscr{P}V$ a $\sigma-$algebra over $V$. A function $\Delta:\mathscr{P}V\rightarrow S(H_{qB})$ is said to be a projective measure if $\Delta$ satisfies the following properties:

\emph{(1)} $\Delta(V)\eqsim I$, $I$ is an identity operator.

\emph{(2)} $\Delta(\bigcup_iA_i)\eqsim\Sigma_i\Delta(A_i)$, for any pairwise disjoint and countable sequence $A_1,A_2,\cdots$ in $V$.

We call the triple $<V,\mathscr{P}V,\Delta>$ is a projective measurable space.
\end{definition}
In accordance with this definition, once a classical valuation is given, a corresponding projective measurement operator will be gained. We will regard a classical valuation as a measurement outcome. If a quantum system is at a quantum state $\rho$, then the probability of the measurement outcome $U$ is $p=tr(\Delta(U)(\rho))$.

A projective measurable space is different from the probabilistic measurable space in\cite{[18]}, where the probabilities are replaced by the super-operators. In fact, a projective measurable space is also called a super-operator valued measure space (or the super-operator valued distributions) in\cite{[25]}. For more details, we refer to the paper\cite{[25]}. In the Section 3, we will use this projective measurable space to propose a quantum operator structure which interprets the semantics of our logic.

\section{Exogenous quantum operator logic}
In this section, we will design an exogenous quantum logic based on density operators from aspects of syntax and semantics, called exogenous quantum operator logic(\textbf{EQOL}). We will use this logic to model and reason about open quantum systems.

\subsection{Syntax of \textbf{EQOL}}

Given a finite set of qubit symbols $qB$, the syntax of \textbf{EQOL} consists of classical formulae, operator terms and quantum operator formulae. We will discuss it in detail below.
\begin{enumerate}
\item[(1)] classical formulae: $\alpha::=\perp\mid qb\mid\alpha\rightarrow\alpha.$
\item[(2)] operator terms: $t::=0\mid I\mid x\mid\int\alpha\mid T^G_A\mid t+t\mid tt\mid t\otimes t$.
\item[(3)] quantum operator formulae: $\gamma::=t\leq t\mid[G]\mid\Perp\mid\gamma\sqsupset\gamma.$
\end{enumerate}

The first syntax part is the classical formulae. Just as \textbf{EQPL}, the classical formulae are the original language of our logic which guides the design of operator terms and quantum operator formulae. We call them the original formulae which are built from qubit symbols in qB by using classical connectives($\neg,\rightarrow$) and falsum $\perp$. As usual, other classical connectives like $\vee,\wedge,\leftrightarrow,\top$ are defined. A collection of all classical formulae is denoted by $\Gamma_C$.

The second syntax part is the operator terms. In the syntax of \textbf{EQPL}, the term language is used to interpret amplitudes, probabilities in real closed fields, and is a core part of \textbf{EQPL}. Instead, we denote the term language in our logic(\textbf{EQOL}) which is interpreted in the super-operators $S(H_{qB})$. Hence, we call them the operator terms. The null operator $0$ and the identity operator $I$ are two constant operators in $S(H_{qB})$. A set of variables $X=\{x_k|k\in N\}$ is interpreted in $S(H_{qB})$, and each $x_k$ is an operator variable term. The operator terms $\int\alpha$ and $T^G_A$ are called the probability operator term, the projective measurement operator term. These two operator terms will be explained in detail in the last part of this subsection. Moveover, we also give the operations between the operator terms, including the addition operator term $t+t$, the composition operator term $tt$, and the tensor product operator term $t\otimes t$. We write a collection of all operator terms as $Term$.

The third syntax part is the quantum operator formulae. We call them the global formulae which are recursively built from quantum operator comparison proposition $t\leq t$ and quantum sub-system [G]($G\subseteq qB$) using the connectives $\sqsupset$ and $\Perp$. We call these two connectives quantum implication and quantum falsum. If a quantum operator formula is just a quantum operator comparison proposition or a sub-system, then we call them quantum operator atomic propositions. We denote the collection of all quantum operator atomic propositions by $qAtom=\{t\leq t,[G]\}$, and write a collection of all quantum operator formulae as $\Gamma_Q$. Also, if there no any operator term variables contained in a given quantum operator formula, then we call it a quantum operator closed formula.

As a supplementary explanation, we need to give the meanings about the probability operator term, the projective measurement operator term and several quantum operator formulae.

The probability operator term $\int\alpha$ is a projective measurement operator on a subspace spanned by a standard computational basis which is a collection of measurement outcomes that make classical formula $\alpha$ true. It is also denoted by$P_{\llbracket\alpha\rrbracket}$where $\llbracket\alpha\rrbracket$ is a set of valuations which make classical formula $\alpha$ true. Given a density operator $\rho$, then $tr((\int\alpha)(\rho))$ denotes a probability that the classical formula $\alpha$ holds for the outcomes occurring in a projective measurement $P_{\llbracket\alpha\rrbracket}$. Please note that there is also a probability term $\int\alpha$ in \textbf{EQPL}. But the difference here is that the latter is interpreted in the real closed field. However, it also denotes a probability that the classical formula $\alpha$ holds for the outcomes occurring in measurement. In this sense, these two terms have the same meanings.

We define the projective measurement operator term $T_A^G$ as follows: $$T_A^G:=P_{(\wedge A)_G}\otimes I_{qB\setminus G},$$ where $A\subseteq G,G\subseteq qB$. If $G=qB$, then we denote $T^{qB}_A$ by $T_A$.

Given $A$ be a subset of $G$, we also define $$\wedge A:=\mathop{\wedge}\limits_{qb_k\in A}qb_k\wedge\mathop{\wedge}\limits_{qb_k\in G\setminus A}\neg qb_k\wedge 2^{qB\setminus G}\equiv
\mathop{\wedge}\limits_{qb_k\in A}qb_k\wedge\mathop{\wedge}\limits_{qb_k\in G\setminus A}\neg qb_k\wedge(\mathop{\vee}\limits_{\bar{x}}
\mathop{\wedge}\limits_{qb_k\in qB\setminus G}(qb_k)^{x_k}),$$ where $(qb_k)^0=\neg qb_k, (qb_k)^1=qb_k$, $\bar{x}=\{x_1x_2\cdots x_{|qB\setminus G|}|x_i\in\{0,1\}\}$, $|qB\setminus G|$ the cardinality of $qB\setminus G$.

If $A$ is restricted to $G$, then we denote $\mathop{\wedge}\limits_{qb_k\in A}qb_k\wedge\mathop{\wedge}\limits_{qb_k\in G\setminus A}\neg qb_k$ by $(\wedge A)_G$, that is, $$(\wedge A)_G=\mathop{\wedge}\limits_{qb_k\in A}qb_k\wedge\mathop{\wedge}\limits_{qb_k\in G\setminus A}\neg qb_k.$$ In particular, we have $(\wedge A)_{qB}=\mathop{\wedge}\limits_{qb_k\in A}qb_k\wedge\mathop{\wedge}\limits_{qb_k\in qB\setminus A}\neg qb_k$.

The $P_{(\wedge A)_G}$ in $T_A^G$ denotes a projective measurement operator on the subspace of Hilbert spanned by a standard computational basis which makes the classical formula $\wedge A$ true, that is, $P_{(\wedge A)_G}=\int(\wedge A)_G$. Meanwhile, the $I_{qB\setminus G}$ in $T_A^G$ is an identity operator restricted to $qB\setminus G$. Furthermore, $\forall A\subseteq G$, we have $$\mathop{\sum}\limits_{A\subseteq G}T_A^G=
\mathop{\sum}\limits_{A\subseteq G}P_{(\wedge A)_G}\otimes I_{qB\setminus G}=I_{G}\otimes I_{qB\setminus G}=I_G.$$

\begin{example}
Let $qB=\{qb_1,qb_2,qb_3,qb_4,qb_5\}$, $G=\{qb_1,qb_2,qb_3,qb_4\}$, and $A=\{qb_1,qb_2\}$, then

\emph{(1)} $(\wedge A)_G=qb_1\wedge qb_2\wedge\neg qb_3\wedge\neg qb_4\equiv(qb_1\wedge qb_2\wedge\neg qb_3\wedge\neg qb_4\wedge\neg qb_5)\vee
(qb_1\wedge qb_2\wedge\neg qb_3\wedge\neg qb_4\wedge\ qb_5)=(11000)\vee(11001)$.

\emph{(2)} $(\wedge A)_{qB}=(qb_1\wedge qb_2\wedge\neg qb_3\wedge\neg qb_4\wedge\neg qb_5)=11000$.

Assume that the Hilbert subspace $H_A$ is spanned by the set $A$, that is, $H_{A}=span\{|1100\rangle\}$, then

\emph{(3)} $T^G_A=P_{(\wedge A)_G}\otimes I_{qB\setminus G}=P_{1100}\otimes I_{qb_5}=P_{1100}\otimes(|0\rangle\langle 0|+|1\rangle\langle 1|)$.

\emph{(4)} $T_A=P_{11000}$.
\end{example}
Based on the above definition, the operator term $T_A^G$ is essentially a projective measurement operator. Similarly, it corresponds to the term $|T\rangle_{GA}$ of \textbf{EQPL}. The latter is called the logical amplitude $\nu_{GA}$\cite{[18]}. Assume that a quantum state is $|\psi\rangle$, $v^G_A$ is a classical valuation which assigns true to elements of $A$ and false to elements of $G\setminus A$, then $||T\rangle_{GA}|=|\langle v^G_A|\psi\rangle|$ denotes a probability that the outcome is $v^G_A$ after the projective measurement. Similarly, the former denotes that when a quantum state $\rho$ is given, $tr(T_A^G(\rho))$ is a probability that the outcome is $(\wedge A)_G$ after the projective measurement $T_A^G$.

The quantum sub-system $[G]$ is the same with that of \textbf{EQPL}. It denotes that $G$ is an isolated and non-entanglement quantum sub-system. The quantum operator comparison proposition $t_1\leq t_2$ denotes that if a density operator $\rho$ is given, then the probability of the measurement outcomes performing $t_1$ at the state $\rho$ is not greater that that of performing $t_2$. It is different from that of \textbf{EQPL}. The latter is a comparison between two numbers from the real closed field. Moreover, it is also different from (1) of Definition 2.1. The latter is a comparison between the two super-operators at the state $\rho$. But, the former is a logical formula which need to be interpreted. In other words, given a quantum state $\rho$, only if after interpretation in $S(H_{qB})$, then $t_1\leq t_2$ is a comparison between the two super-operators, that is, $t_1\lesssim_{\rho} t_2$.

\begin{example}
Let $qB=\{qb_1,qb_2\}$, $\alpha=qb_1\wedge qb_2$, $\beta=\neg qb_1\wedge\neg qb_2$, then $\llbracket\alpha\rrbracket=\{11\}$, $\llbracket\beta\rrbracket=\{00\}$. We consider the following quantum operator comparison proposition
\begin{center}
$(\int\alpha)\leq(\int\beta)$ or $(P_{\llbracket\alpha\rrbracket}\leq P_{\llbracket\beta\rrbracket}).$
\end{center}
If given a quantum state $\rho=0.7|00\rangle\langle 00|+0.3|11\rangle\langle 11|$, then we have
\begin{center}
$tr((\int\alpha)\rho(\int\alpha)^{\dagger})\leq tr((\int\beta)\rho(\int\beta)^{\dagger})$ i.e., $P_{\llbracket\alpha\rrbracket}\lesssim_{\rho} P_{\llbracket\beta\rrbracket}.$
\end{center}
It implies that $(\int\alpha)\leq(\int\beta)$ is true at the state $\rho$.

If a quantum state $\rho=0.3|00\rangle\langle 00|+0.7|11\rangle\langle 11|$, then we have $P_{\llbracket\alpha\rrbracket}\lnsim_{\rho} P_{\llbracket\beta\rrbracket}$ which implies that $(\int\alpha)\leq(\int\beta)$ is false at the state $\rho$.
\end{example}
\begin{definition}
Considering an exogenous quantum operator logic, we define a sub-language of \textbf{EQOL} by

$a:=0\mid I\mid x\mid a+a\mid aa\mid a\otimes a$, where $a\in S(H_{qB})$;

$\kappa:=a\leq a\mid\Perp\mid\kappa\sqsupset\kappa$.
\end{definition}
We denoted this sub-language by \textbf{sub-EQOL}. We call the terms of \textbf{sub-EQOL} as quantum operator analytical terms, the quantum operator formulae of \textbf{sub-EQOL} as quantum operator analytical formulae. This \textbf{sub-EQOL} is useful to study the completeness of \textbf{EQOL}.

\subsection{Semantic of \textbf{EQOL}}

Given a density operator $\rho$, the language of \textbf{EQOL} needs to be interpreted. The semantics of \textbf{EQOL} consist of three components: the valuations of classical formulae, the interpretations of quantum operator terms, the satisfactions of quantum operator formulae.

1. The valuation of classical formulae.

Assume that $\alpha$ is a classical formula and $v$ is a valuation in $2^{qB}$, we denote $v\models\alpha$ if $v$ satisfies $\alpha$, and write $\llbracket\alpha\rrbracket=\{v\in 2^{qB}|v\models\alpha\}$ as a valuation set of $\alpha$.

2. The interpretation of quantum operator terms.

According to quantum operator terms, we will propose their semantics from two aspects: the operator interpretations and the probability interpretations.

Firstly, we need to define an assignment function $\sigma:X\rightarrow S(H_{qB})$ where $X$ is a collection of all operator variables, and an operator interpretation function of terms: $$[\cdot]:Term\rightarrow S(H_{qB}).$$ Obviously, the function $\sigma$ is the restriction of $[\cdot]$ to $X$. For simplicity, in most cases, we will not consider the operators variable terms and always omit the symbol of the assignment function $'\sigma'$.

We present the operator interpretations as follows.

(1) $[x]\eqsim\sigma(x)$; (2) $[0]\eqsim 0$; (3) $[I]\eqsim I$; (4) $[\int\alpha]\eqsim P_{\llbracket\alpha\rrbracket}\eqsim\mathop{\sum}\limits_{v\in\llbracket\alpha\rrbracket}P_v$; (5) $[T_A^G]\eqsim P_{(\wedge A)_G}\otimes I_{qB\setminus G}$, where $A\subseteq G,G\subseteq qB$; (6) $[t_1+t_2]\eqsim [t_1]+[t_2]$; (7) $[t_1t_2]\eqsim [t_1]\cdot[t_2]$; (8) $[t_1\otimes t_2]\eqsim[t_1]\otimes [t_2]$.

Secondly, we will present the probability interpretations of terms in the following. We need to define a quantum operator structure $M$ over a projective measurable space.

In accordance with\cite{[18]}, we introduce a definition about the component sub-systems. Assume that $S$ be a partition of $qB$, then $Alg(S)$ is defined by $\{\bigcup_iA_i:\{A_i\}\subseteq S\}$, that is, it is a collection of all unions of sets in the partition(including the emptyset). Each $G\in Alg(S)$ models the qubits of the component systems. $Alg(S)$ satisfies the following properties:

(1) $\phi,qB\in Alg(S)$;

(2) If $A\in Alg(S)$, then $qB\setminus A\in Alg(S)$;

(3) If $A_1,A_2\in Alg(S)$, then $A_1\bigcup A_2\in Alg(S)$.

\begin{definition}
A quantum operator structure over a projective measurable space $<V,\mathscr{P}V,\Delta>$ is a five-tuple $M=<S(H_{qB}),V,S,\rho,\mu>$, where:

\emph{(1)} $V$ is a non-empty subset of $2^{qB}$;

\emph{(2)} $S$ is a partition of $qB$;

\emph{(3)} $\rho$ is a density operator on $H_{qB}$;

\emph{(4) }$P_{v}=0$, $v\not\in V$;

\emph{(5)} $\mu:=\{T_{A}^G|A\subseteq G, G\subseteq {qB}\}$, where $T_A^G\eqsim P_{(\wedge A)_G}\otimes I_{qB\setminus G}$, if $G\in Alg(S)$.
\end{definition}

\begin{example}
We consider the following quantum operator structure $$M=<S(H_{qB}),V,S,\rho,\mu>,$$ where $qB=\{qb_1,qb_2\}$, $V=\{00,01,10,11\}$, $S=\{\emptyset, qB\}$, $\rho=0.4|00\rangle\langle 00|+0.6|11\rangle\langle 11|$.

We can obtain $$Alg(S)=\{\emptyset,qB\},$$ and have $\mu=\{T_{\emptyset}^{\emptyset},T_{\emptyset}^{qB}, T_{qB}^{qB}\}$.
\end{example}

Given a set of density operators $\{\rho_s,s\in S\}$, where $\rho_s$ is a density operator on $H_s$, we extend $\rho$ to $Alg(S)$ as follows.

(1) $\rho_{\emptyset}=I$;

(2) $\rho_{A_1\cup A_2\cdots\cup A_n}=\mathop{\otimes}\limits_{i=1}^{n}\rho_{A_i}$.

Let a quantum operator structure $M=<S(H_{qB}),V,S,\rho,\mu>$ be described by a quantum operator structure $<V,\mathscr{P}V,\Delta>$, where for any $U\subseteq V$, $$\Delta(U)\eqsim\Sigma_{v\in U}P_v\in S(H_{qB}).$$

Given a quantum operator structure $M=<S(H_{qB}),V,S,\rho,\mu>$, we also need to define a probability interpretation function of terms over $M$: $$\llbracket\cdot\rrbracket _M: Term\rightarrow [0,1],$$ for $t\in Term$, $\llbracket t\rrbracket _M=tr([t](\rho))$. This function acts the operator interpretation about $t$ first and calculates the trace at the state $\rho$ later. For simplicity, we always omit the symbol $'[ ]'$.

Based on the above analysis, we recursively define the interpretations of quantum operator terms as follows.

(1) $\llbracket x\rrbracket _M=tr(\sigma(x)(\rho))$.

(2) $\llbracket 0\rrbracket_M=tr(0(\rho))=0$.

(3) $\llbracket I\rrbracket_M=tr(I(\rho))=1$.

(4) $\llbracket\int\alpha\rrbracket_M=tr(P_{\llbracket\alpha\rrbracket}(\rho))=\mathop{\sum}\limits_{v\in\llbracket\alpha\rrbracket}tr(P_v(\rho))$.

(5) $\llbracket T_A^G\rrbracket_M=tr((P_{(\wedge A)_G}\otimes I_{qB\setminus G})(\rho))$, where $A\subseteq G,G\subseteq qB$.

(6) $\llbracket t_1+t_2\rrbracket_M=\llbracket t_1\rrbracket_M+\llbracket t_2\rrbracket_M=
tr(t_1(\rho))+tr(t_2(\rho))$.

(7) $\llbracket t_1t_2\rrbracket_M=tr(t_1(t_2(\rho)))$.

(8) $\llbracket t_1\otimes t_2\rrbracket_M=tr((t_1\otimes t_2)(\rho))=
tr(t_1(\rho_1))tr(t_2(\rho_2))$, where $\rho=\rho_1\otimes\rho_2$.

\begin{example}
Let a quantum operator structure be given as in Example 3.3, then we have

\emph{(1)} $\llbracket\int\neg qb_1\vee\neg qb_2\rrbracket_M=tr((P_{00}+P_{01}+P_{10})(\rho))=0.4$;

\emph{(2)} $\llbracket T_{qB}^{qB}\rrbracket_M=tr(P_{11}(\rho))=0.6$;

\emph{(3)} $\llbracket T_{\emptyset}^{qB}T_{qB}^{qB}\rrbracket_M=tr(P_{00}(P_{11}(\rho)))=0$.
\end{example}

The difference from the interpretations of terms in \textbf{EQPL} is that we interpret the operator terms as the super-operators. Meanwhile, if a quantum operator structure is given, then we also present their probability interpretations. For example, $\int\alpha$ is interpreted as a probability that the classical formula $\alpha$ holds for an outcome of a projective measurement, and $T_A^G$ is interpreted as a probability that the outcome is $(\wedge A)_G$ after measurement.

3. The satisfactions of quantum operator formulae

Let $M=<S(H_{qB}),V,S,\rho,\mu>$ be a quantum operator structure, then we recursively define the satisfactions of quantum operator formulas as follows.

(1) $M\models t_1\leq t_2$ if and only if $[t_1]\lesssim_{\rho}[t_2]$, or $\llbracket t_1\rrbracket_M\leq\llbracket t_2\rrbracket_M$.

(2) $M\not\models\Perp$.

(3) $M\models [G]$ if and only if $G\in Alg(S)$.

(4) $M\models\gamma_1\sqsupset\gamma_2$ if and only if $M\not\models\gamma_1$, or
$M\models\gamma_2$.

The satisfaction of quantum operator comparison proposition denotes that a given quantum operator structure $M$ satisfies $t_1\leq t_2$ if and only if the probability of measurement outcomes of performing $t_1$ is not greater than that of performing $t_2$ on $M$. The satisfaction of $[G]$ denotes that a given quantum operator structure $M$ satisfies $[G]$ if and only if $G$ is an isolated quantum sub-system. The satisfaction of $\gamma_1\sqsupset\gamma_2$ is similar with that of the classical implication formula, that is, a given quantum operator structure $M$ satisfies the $\gamma_1\sqsupset\gamma_2$ if and only if either $M$ does not satisfy $\gamma_1$ or $M$ satisfies $\gamma_2$.

\subsection{Axioms and rules}

In order to enrich exogenous quantum operator logic, we further define several other quantum operator formulae and connectives, and give their shorthand forms as follows.

(1) quantum operator negation: $\boxminus\gamma$ for $\gamma\sqsupset\Perp$.

(2) quantum operator disjunction: $\gamma_1\sqcup\gamma_2$ for $(\boxminus\gamma_1)\sqsupset\gamma_2$.

(3) quantum operator conjunction: $\gamma_1\sqcap\gamma_2$ for $\boxminus(\boxminus\gamma_1\sqcup\boxminus\gamma_2)$.

(4) operator terms equivalence: $t_1=t_2$ for $(t_1\leq t_2)\sqcap(t_2\leq t_1)$.

(5) quantum operator equivalence: $\gamma_1\equiv\gamma_2$ for $ (\gamma_1\sqsupset\gamma_2)\sqcap(\gamma_2\sqsupset\gamma_1)$.

(6) $t_1$ less than $t_2$: $t_1<t_2$ for $(t_1\leq t_2)\sqcap\boxminus(t_2\leq t_1)$.

Given a collection of all classical propositional variables, denoted by $P$, we write $f:P\rightarrow qAtom$ as a mapping from $P$ to quantum operator axiomic propositions $qAtom$. Then, we recursively extend $f$ to a homomorphic mapping from classical formulae $\Gamma_C$ to quantum operator formulae $\Gamma_Q$, that is, for any classical formulae $\alpha_1, \alpha_2$, $f(\neg\alpha)=\boxminus f(\alpha)$, $f(\alpha_1\rightarrow\alpha_2)=f(\alpha_1)\sqsupset f(\alpha_2)$. Meanwhile, if $\alpha$ is a classical formula, we write $\alpha_{qB}f$ as a quantum operator formula under a homomorphic mapping $f$.

\begin{definition}
Let $\top$ be a classical proposition tautology, $\top_{qB}f$ is defined as a quantum operator tautology, denoted by $\Top$.
\end{definition}

We will present an axiomatization system of \textbf{EQOL} which contains 12 axioms and 2 rules as follows. Every axiom or rule will be named. For example, the first axiom is named by QTaut.

(1) 12 axioms

$[QTaut]$  $\vdash\gamma$, for each quantum operator tautology $\gamma$.

$[RCF]$  $\vdash k\{\vec{x}/\vec{t}\}$, where $k$ is a quantum operator tautology, $\vec{x}$ and $\vec{t}$ are sequences of variables and terms. The $k\{\vec{x}/\vec{t}\}$ implies that it is obtained by replacing all occurrences of $x_i$ by $t_i$.

$[Unit]$  $\vdash[G]\sqsupset(\mathop{\sum}\limits_{A\subseteq G}T_A^G=I)$, where $G\subseteq qB$, especially, $\vdash \mathop{\sum}\limits_{A\subseteq qB}T_A=I$.

$[CTaut]$  $\vdash\int\alpha=I$, for each classical tautology $\alpha$.

$[Mesh\emptyset]$  $\vdash\int\perp=O$.

$[FAdd]$  $\vdash(\int\alpha_1\wedge\alpha_2=O)\sqsupset(\int\alpha_1\vee\alpha_2=\int\alpha_1+\int\alpha_2)$.

$[Mon]$  $\vdash\int\alpha_1\rightarrow\alpha_2\sqsupset(\int\alpha_1\leq\int\alpha_2)$.

$[Prob]$  $\vdash\int\wedge A=T_A$, or $\vdash\int\alpha=\mathop{\sum}\limits_{A\in\llbracket\alpha\rrbracket}T_A$.

$[MO1]$  $\vdash([G_1]\sqcap[G_2])\sqsupset(T^{G_1\cup G_2}_{A_1\cup A_2}=T^{G_1}_{A_1}\otimes T^{G_2}_{A_2}$), where $G_1\cap G_2=\emptyset,A_1\subseteq G_1, A_2\subseteq G_2$.

$[Sub\emptyset]$  $\vdash[\emptyset]$.

$[Sub\cup]$  $\vdash[G_1]\sqsupset([G_2]\sqsupset[G_1\sqcup G_2]) $.

$[Sub\setminus]$  $\vdash[G]\equiv[qB\setminus G] $.

(2) 2 rules

$[CMP]$  $\alpha_1,\alpha_1\rightarrow\alpha_2\vdash\alpha_2$.

$[QMP]$  $\gamma_1,\gamma_1\sqsupset\gamma_2\vdash\gamma_2$.

The $Sub\emptyset$, $Sub\cup$, $Sub\setminus$ and $CMP$ are the same with those of \textbf{EQPL}\cite{[18]} in the sense of syntax and semantic. Through $QTaut$, $RCF$, $Mesh\emptyset$, $FAdd$ and $QMP$ have the same forms with those of \textbf{EQPL}, they are proposed in the sense of operator. The $Unit$, $CTaut$, $Mon$, $Prob$ preserve the semantic equivalence from those of \textbf{EQPL}. The $MO1$ mainly concerns the tensor product between projective measurement operators.

\subsection{The satisfiability problem of exogenous quantum operator formulae}

As for the satisfiability problem of $\Gamma_Q$, we only concern about quantum operator closed formulae in this subsection. Hence, we need no assignment functions. Given any quantum operator formulae $\gamma,\gamma_1,\gamma_2\in\Gamma_Q$, the length $|\gamma|$ of a quantum operator formula $\gamma$ is defined recursively as follows: (1) $|\gamma|=0$; (2) $|\boxminus\gamma|=|\gamma|+1$; (3) $|\gamma_1\sqsupset\gamma_2|=max(|\gamma_1|,|\gamma_2|)+1$. If there is an algorithm to decide whether a given quantum operator structure satisfies a quantum operator closed formula, then we have:

\begin{theorem} Given a quantum operator structure $M$, $\gamma$ is a quantum operator closed formula,
then we need an $O(2^{4n}|\gamma|)$ time to decide whether $M$ satisfies $\gamma$.
\end{theorem}
\noindent {\bf Proof.}
We assume that all basic arithmetical operations take unit time. As to every quantum closed operator formula $\gamma$, its length of $\gamma$ is $|\gamma|$. Let $|qB|=n$, every quantum density operator over $qB$ is represented by a a $2^n\times 2^n$ matrix. So the addition, the subtraction and the multiplication between matrixes respectively takes $O(2^{2n})$,$O(2^{2n})$,$O(2^{3n})$ time. As to the probability operator term $\int\alpha$, the evaluation takes $O(2^{4n})$ time as we have to compute the set of valuation $2^{qB}$ and $\llbracket\int\alpha\rrbracket_M=tr(P_{\llbracket\alpha\rrbracket}(\rho))$, where we require $O(2^{n})$ time corresponding to traveling throughout all the valuations satisfying $\alpha$ and consume $O(2^{3n})$ time about $tr(P_{v}(\rho))$ for every valuation. By reason of $\llbracket T_A^G\rrbracket_M=tr(T_A^G(\rho))$, we take $O(2^{3n})$ time to interpret $T^G_A$. According to interpretation of the quantum operator terms $t_1+t_2$, $t_1t_2$ and $t_1\otimes t_2$, computing these terms respectively require $O(2^{n})$, $O(2^{3n})$, $O(2^{4n})$ time. After the time complexities of these operator terms are obtained, the remaining computation about quantum operator formula $\gamma$ takes at most $O(|\gamma|)$ time. Hence, the total time algorithm to decide if $M$ satisfies $\gamma$ is $O(2^{4n})$. \qed

\section{Soundness}
As an axiomatization system, a logic is sound which implies that if $\Gamma\vdash\gamma$, then $\Gamma\models\gamma$. The proof of soundness suffices to show that if $\gamma$ is an axiom, then any given semantic structure(model) $M$ satisfies $\gamma$. That is, every axiom is valid. In this section, we will show that exogenous quantum operator logic is sound.

\begin{lemma}
Let $\alpha$ be a classical tautology and $f$ be a homomophic from $\Gamma_C$ to $\Gamma_Q$, for any quantum operator tautology $\alpha_{qB}f$, if $\vdash\alpha_{qB}f$, then $\models\alpha_{qB}f$.
\end{lemma}

\noindent {\bf Proof.}
Given a collection of all classical propositional variables $P$, a quantum operator structure $M=<S(H_{qB}),V,$ $S,\rho,\mu>$,  we define a valuation function $v'$ over $P$, i.e., $v':P\rightarrow \{0,1\}$ such that $\forall p\in P$,
$$v'(p)=\left\{
          \begin{array}{ll}
            1, & \hbox{$M\models f(p)$,} \\
            0, & \hbox{otherwise.}
          \end{array}
        \right.
$$

For any classical propositional formula $\alpha$, it is easy to show that the following proposition holds by using induction on the structure of the formula $\alpha$,
\begin{center}
$v'\models\alpha$ if and only if $M\models\alpha_{qB} f$.
\end{center}

(1) If $\alpha$ is a propositional formula $\bot$(falsum), we have $v'(\bot)\equiv0$ if and only if $M\not\models\bot_{qB}f$.

(2) For any $v\in V$, if $\alpha$ is $\alpha_1\rightarrow\alpha_2$, then $v$ doesn't satisfy $\alpha_1$ or $v$ satisfies $\alpha_2$. If $v\models\alpha_2$, we show by using induction on the structure of the formula $\alpha_2$ that there is a quantum operator structure $M$ such that $M\models(\alpha_2)_{qB}f$. If $v\not\models\alpha_1$, then there is not a quantum operator structure $M$ such that $M\models(\alpha_1)_{qB}f$. So for any $M$, $M\not\models(\alpha_1)_{qB}f$ or $M\models(\alpha_2)_{qB}f$, that is, $M\models(\alpha_1)_{qB}f\sqsupset(\alpha_2)_{qB}f$. Let $\alpha_{qB}f:=(\alpha_1)_{qB}f\sqsupset(\alpha_2)_{qB}f$, then we obtain $M\models\alpha_{qB}f$. Therefore, for any $v\in V$, if $v\models\alpha$, then there is a quantum operator structure $M$ such that $M\models(\alpha)_{qB}f$. Conversely, according to the definition of $v'$, we have $v'\models\alpha$. From what we have showed above, $v'\models\alpha$ if and only if $M\models\alpha_{qB} f$. In particular, if $\alpha$ is a classical propositional tautology $\top$, then $\models\top$ if and only if $\models\top_{qB} f$. Since the classical propositional logic is sound, we obtain that if $\alpha$ is an axiom($\vdash\alpha$), then $\alpha$ is valid($\models\alpha$). Hence $\alpha_{qB} f$ is also valid, i.e., $\models\alpha_{qB} f$.

Under the homomorphic mapping, the quantum operator formula $\top_{qB}f$(the quantum operator formula $\bot_{qB}f$) corresponds for tautology $\top$ (contradiction $\bot$). We also denote $\top_{qB}f$ by $\Top$. In accordance with Lemma 4.1, we get the following conclusion: $\bot_{qB}f\equiv \Perp$, $\boxminus\Perp\equiv\Top$.\qed

\begin{lemma} The axioms are valid. That is, let $\gamma$ be a quantum operator formula, if $\vdash\gamma$, then $\models\gamma$.
\end{lemma}
\noindent {\bf Proof.}
Given a quantum operator structure $M$, we get

($\textbf{Unit}$) If $\vdash([G])\sqsupset(\mathop{\sum}\limits_{A\subseteq G}T_A^G=I)$, then $\models([G])\sqsupset(\mathop{\sum}\limits_{A\subseteq G}T_A^G=I)$.

Assume that $M\models[G]$, then $G\in Alg(S)$ and $qB\setminus G\in Alg(S)$ which imply that $G$ and $qB\setminus G$ are distinguishable and non-entangled. Moreover, we have

\qquad\qquad$\llbracket \mathop{\sum}\limits_{A\subseteq G}T_A^G\rrbracket_{M}=\mathop{\sum}\limits_{A\subseteq G}tr(T_A^G(\rho))=\mathop{\sum}\limits_{A\subseteq G}tr((P_{(\wedge A)_G}\otimes I_{qB\setminus G})(\rho))$

\qquad\qquad\qquad\qquad\ \ $=\mathop{\sum}\limits_{A\subseteq G}tr((P_{(\wedge A)_G}\otimes I_{qB\setminus G})(\rho_G\otimes\rho_{qB\setminus G}))=\mathop{\sum}\limits_{A\subseteq G}tr((P_{(\wedge A)_G}(\rho_G))=1$.
\\
Hence, $\models([G])\sqsupset(\mathop{\sum}\limits_{A\subseteq G}T_A^G=I)$.

($\textbf{Mesh}\emptyset$) If $\vdash\int\perp=0$, then $\models\int\perp=0$.

Using the definition $\llbracket\int\perp\rrbracket_{M}=\mathop{\sum}\limits_{v\in\llbracket\perp\rrbracket}
tr(P_v\rho)=\mathop{\sum}\limits_{\emptyset}P_{\emptyset}\rho=tr(0\rho)=0$ and $\llbracket 0\rrbracket_{M}=0$, then we obtain $\models\int\perp=0$.

($\textbf{FAdd}$) If $\vdash(\int\alpha_1\wedge\alpha_2=0)\sqsupset(\int\alpha_1\vee\alpha_2
=\int\alpha_1+\int\alpha_2)$, then $\models(\int\alpha_1\wedge\alpha_2=0)\sqsupset(\int\alpha_1\vee\alpha_2
=\int\alpha_1+\int\alpha_2)$.

Assume that $M\models\int\alpha_1\wedge\alpha_2=0$,
then $\alpha_1\wedge\alpha_2\equiv\perp$ and so
$\llbracket\alpha_1\wedge\alpha_2\rrbracket=\emptyset$, hence
$\llbracket\alpha_1\rrbracket\cap\llbracket\alpha_2\rrbracket=\emptyset$. Thus,
if $\forall v\in\llbracket\alpha_1\rrbracket$, then $v\not\in\llbracket\alpha_2\rrbracket$, or if
$\forall v\in\llbracket\alpha_2\rrbracket$, then $v\not\in\llbracket\alpha_1\rrbracket$.

Again, $\llbracket\int\alpha_1\vee\alpha_2\rrbracket_{M}=\mathop{\sum}
\limits_{v\in\llbracket\alpha_1\vee\alpha_2\rrbracket}tr(P_v\rho)=(\mathop{\sum}
\limits_{v\in\llbracket\alpha_1\rrbracket}+\mathop{\sum}
\limits_{v\in\llbracket\alpha_2\rrbracket})tr(P_v\rho)$

\qquad\qquad\qquad\qquad\ \ $=\mathop{\sum}
\limits_{v\in\llbracket\alpha_1\rrbracket}tr(P_v\rho)+\mathop{\sum}
\limits_{v\in\llbracket\alpha_2\rrbracket}tr(P_v\rho)=
\llbracket\int\alpha_1\rrbracket_{M}+
\llbracket\int\alpha_2\rrbracket_{M}$.
\\Therefore, $\int\alpha_1\vee\alpha_2=\int\alpha_1+\int\alpha_2$. We get

\qquad\qquad\qquad\qquad\ \ $M\models(\int\alpha_1\wedge\alpha_2
=0)\sqsupset(\int\alpha_1\vee\alpha_2
=\int\alpha_1+\int\alpha_2)$.
\\
Since the quantum operator structure $M$ is arbitrary, we get

\qquad\qquad\qquad\qquad\ \ $\models(\int\alpha_1\wedge\alpha_2=0)\sqsupset(\int\alpha_1\vee\alpha_2
=\int\alpha_1+\int\alpha_2)$.

($\textbf{Mon}$) If $\vdash(\int\alpha_1\rightarrow\alpha_2)\sqsupset(\int\alpha_1\leq\int\alpha_2)$, then $\models(\int\alpha_1\rightarrow\alpha_2)\sqsupset(\int\alpha_1
\leq\int\alpha_2)$.

Assume that $M\models \int\alpha_1\rightarrow\alpha_2$, then $M\models\int\alpha_1\rightarrow\alpha_2\equiv I$.
So, we get $\llbracket\alpha_1\rightarrow\alpha_2\rrbracket=2^{qB}$. This implies that $(\alpha_1\rightarrow\alpha_2)\equiv\top$ or
$\llbracket\alpha_1\rrbracket\subseteq\llbracket\alpha_2\rrbracket$. Hence
$\mathop{\sum}\limits_{v\in\llbracket\alpha_1\rrbracket}tr(P_v\rho)\leq\mathop{\sum}
\limits_{v\in\llbracket\alpha_2\rrbracket}tr(P_v\rho)$ and so
$\int\alpha_1\leq\int\alpha_2$. Therefore,
$\models(\int\alpha_1\rightarrow\alpha_2)\sqsupset(\int\alpha_1
\leq\int\alpha_2)$.

($\textbf{Prob}$) If $\vdash\int\wedge A=T_A$, then $\models\int\wedge A=T_A$.

Assume that $qB=\{qb_1,qb_2,\cdots,qb_n\}$ and $A=\{qb'_1,qb'_2,\cdots,qb'_m\}$, $m\leq n$, we get $\wedge A=qb'_1\wedge qb'_2\wedge\cdots\wedge qb'_m\wedge 2^{qB\setminus A}$. Let us take any quantum operator structure $M$ such that

\qquad\qquad$\llbracket\int\wedge A\rrbracket_{M}=
\mathop{\sum}\limits_{v\in\wedge A}tr(P_v\rho)=tr(\mathop{\sum}\limits_{v\in\wedge A}P_v\rho)=tr(\mathop{\sum}\limits_{w\in 2^{qB\setminus A}}P_{qb_1qb_2\cdots qb_nw}\rho)
$

\qquad\qquad\qquad\qquad$=tr((P_{qb_1qb_2\cdots qb_n}\otimes \mathop{\sum}\limits_{w\in 2^{qB\setminus A}}P_w)\rho)
=tr((P_A\otimes I_{2^{qB}\setminus A})\rho)$,

we get $\llbracket T_A\rrbracket_{M}=tr((P_A\otimes I_{2^{qB}/A})\rho)$ and $\models\int\wedge A= T_A$.

($\textbf{MO1}$) If $\vdash([G_1]\sqcap[G_2])\sqsupset (T^{G_1\cup G_2}_{A_1\cup A_2}= T^{G_1}_{A_1}\otimes T^{G_2}_{A_2})$ where $G_1\cap G_2=\emptyset, A_1\subseteq G_1, A_2\subseteq G_2$, then $\models([G_1]\sqcap[G_2])\sqsupset (T^{G_1\cup G_2}_{A_1\cup A_2}= T^{G_1}_{A_1}\otimes T^{G_2}_{A_2})$.

Assume that $M\models[G_1]\sqcap[G_2]$, then we get $M\models[G_1]$ and $M\models[G_2]$, hence $G_1\in Alg(S)$ and $G_2\in Alg(S)$. Again, because of $A_1\subseteq G_1$ and $A_2\subseteq G_2$, then we have $A_1\in Alg(S)$, $A_2\in Alg(S)$ and $A_1\cup A_2\in Alg(S)$. This shows that $G_1$ and $G_2$, $A_1$ and $A_2$ are distinguishable. Hence, we may further compute the following equation.

$\llbracket T^{G_1\cup G_2}_{A_1\cup A_2}\rrbracket_{M}=\mathop{\sum}\limits_{A_1\cup A_2
\subseteq G_1\cup G_2}tr((P_{A_1\cup A_2}\otimes I_{(G_1\cup G_2)\setminus(A_1\cup A_2)})\rho_{G_1\cup G_2})$

\qquad\qquad\ \ $=\mathop{\sum}\limits_{A_1\cup A_2
\subseteq G_1\cup G_2}tr((P_{A_1\cup A_2}\otimes I_{(G_1\cup G_2)\setminus(A_1\cup A_2)})(\rho_{A_1\cup A_2}\otimes\rho_{(G_1\cup G_2)\setminus(A_1\cup A_2)}))$

\qquad\qquad\ \ $=\mathop{\sum}\limits_{A_1\cup A_2
\subseteq G_1\cup G_2}tr((P_{A_1\cup A_2}\rho_{A_1\cup A_2})\otimes(I_{(G_1\cup G_2)\setminus(A_1\cup A_2)}\rho_{(G_1\cup G_2)\setminus(A_1\cup A_2)}))$

\qquad\qquad\ \ $=\mathop{\sum}\limits_{A_1\cup A_2
\subseteq G_1\cup G_2}tr((P_{A_1\cup A_2}\rho_{A_1\cup A_2}))tr((I_{(G_1\cup G_2)\setminus(A_1\cup A_2)}\rho_{(G_1\cup G_2)\setminus(A_1\cup A_2)}))$

\qquad\qquad\ \ $=\mathop{\sum}\limits_{A_1\cup A_2
\subseteq G_1\cup G_2}tr(P_{A_1}\rho_{A_1})tr(P_{A_2}\rho_{A_2})
tr(I_{G_1\setminus A_1}\rho_{G_1\setminus A_1})
tr(I_{G_2\setminus A_2}\rho_{G_2\setminus A_2})$

\qquad\qquad\ \ $=\mathop{\sum}\limits_{A_1\cup A_2
\subseteq G_1\cup G_2}tr(P_{A_1}\rho_{A_1})tr(I_{G_1\setminus A_1}\rho_{G_1\setminus A_1})tr(P_{A_2}\rho_{A_2})
tr(I_{G_2\setminus A_2}\rho_{G_2\setminus A_2})$

\qquad\qquad\ \ $=\mathop{\sum}\limits_{A_1\cup A_2
\subseteq G_1\cup G_2}tr(P_{A_1}\otimes I_{G_1\setminus A_1}\rho_{G_1})tr(P_{A_2}\otimes I_{G_2\setminus A_2}\rho_{G_2})$

\qquad\qquad\ \ $=\mathop{\sum}\limits_{A_1\subseteq G_1}tr(P_{A_1}\otimes I_{G_1\setminus A_1}\rho_{G_1})\mathop{\sum}\limits_{A_2\subseteq G_2}tr(P_{A_2}\otimes I_{G_2\setminus A_2}\rho_{G_2})$.

Using the definition $\llbracket T^{G_1}_{A_1}\rrbracket_{M}\llbracket T^{G_2}_{A_2}\rrbracket_{M}\equiv \mathop{\sum}\limits_{A_1\subseteq G_1}tr(P_{A_1}\otimes I_{G_1\setminus A_1}\rho_{G_1})\mathop{\sum}\limits_{A_2\subseteq G_2}tr(P_{A_2}\otimes I_{G_2\setminus A_2}\rho_{G_2})$, we get $T^{G_1\cup G_2}_{A_1\cup A_2}=T^{G_1}_{A_1}\otimes T^{G_2}_{A_2}$. Hence
$M\models T^{G_1\cup G_2}_{A_1\cup A_2}= T^{G_1}_{A_1}\otimes T^{G_2}_{A_2}$. Since the quantum operator structure $M$ is arbitrary, we get the following result $\models([G_1]\sqcap[G_2])\sqsupset (T^{G_1\cup G_2}_{A_1\cup A_2}\equiv T^{G_1}_{A_1}\otimes T^{G_2}_{A_2})$.\qed

Using Lemma 4.1 and Lemma 4.2, we obtain the following theorem.

\begin{theorem} Assume that any exogenous quantum operator formula $\gamma$, if $\vdash\gamma$, then $\models\gamma$. That is, exogenous quantum operator logic is sound.
\end{theorem}

\section{Completeness}

As an axiomatization system, a logic is complete which implies that if $\Gamma\models\gamma$, then $\Gamma\vdash\gamma$. For an exogenous quantum propositional logic, the proof of its completeness has already been presented in\cite{[13]} which mainly uses the Model Existence Lemma about the consistent exogenous quantum formulae. In this section, a similar techology is proposed to prove the completeness of exogenous quantum operator logic.

\begin{proposition} Every quantum operator formula is a quantum disjunctive normal form.
\end{proposition}
Let $Q\subseteq qAtom$ and $D\subseteq Q$, $(\sqcap_{\mu\in D}\mu)\sqcap(\sqcap_{\mu\in(Q\setminus D)}(\boxminus\mu))$ is said to be a quantum operator molecule formula, denoted by $\sqcap_Q D$, where $D$ and $Q\setminus D$ are the positive and negative part respectively. If $\eta$ is a quantum operator molecule formula, we respectively denote the positive part and the negative part by $\eta^{+},\eta^{-}$. Then, we say that a quantum operator formula is in the disjunctive normal form if it is a disjunctive of quantum operator molecule formulae.

\begin{definition}
A quantum operator formula $\gamma$ is said to be consistent if $\not\vdash\boxminus\gamma$.
\end{definition}
\begin{proposition} Every quantum operator formula is consistent if and only if its quantum operator molecule formulae are at least consistent.
\end{proposition}

\noindent {\bf Proof.}
Without loss of generality, we provide a quantum operator formula $\gamma=\gamma_1\sqcup\gamma_2$. Then $\gamma$ is consistent, if and only if $\gamma_1,\gamma_2$ are at least consistent.

($\Longrightarrow$). Proof by contradiction. In the following, it suffices to show that the quantum disjunction $\gamma$ of two inconsistent quantum operator $\gamma_1$ and $\gamma_2$ is inconsistent. Assume that $\gamma_1$ and $\gamma_2$ are inconsistent, then we have  $\vdash\boxminus\gamma_1$ and $\vdash\boxminus\gamma_2$. Since $(\boxminus\gamma_1)\sqsupset((\boxminus\gamma_2)\sqsupset\boxminus(\gamma_1\sqcup\gamma_2))$, we get $\vdash(\boxminus\gamma_2)\sqsupset\boxminus(\gamma_1\sqcup\gamma_2)$. Hence, $\vdash\boxminus(\gamma_1\sqcup\gamma_2)$. So, $\gamma_1\sqcup\gamma_2$ is inconsistent.

($\Longleftarrow$). Proof by contradiction. In the following, it suffices to show that if $\gamma$ is inconsistent, then $\gamma_1,\gamma_2$ are all inconsistent. Assume that $\gamma$ is inconsistent, then we have $\vdash\boxminus(\gamma_1\sqcup\gamma_2)$. Since $\boxminus(\gamma_1\sqcup\gamma_2)\equiv\boxminus\gamma_1\sqcap\boxminus\gamma_2$, we get $\vdash\boxminus\gamma_1\sqcap\boxminus\gamma_2$. Therefore, $\gamma_1,\gamma_2$ are inconsistent.\qed

According to Proposition 5.2, in order to decide consistency of exogenous quantum operator formulae $\gamma$, we only need  to show that one of its quantum operator molecules is consistent.

Assume that any given quantum operator formula $\gamma$, we can find an equivalent quantum operator formula $\gamma'$ such that $\gamma'$ has no any probability operator term $\int\alpha$. It is followed from Axioms $CTaut$, $Mesh\emptyset$, $FAdd$, $Mon$, $Prob$.

\begin{proposition} Let $\eta$ be a quantum operator molecule formula, there is a molecule formula $\eta'$ such that $\eta'$ has no any probability operator term $\int\alpha$ and $\vdash\eta\equiv\eta'$.
\end{proposition}

\noindent {\bf Proof.}
Without loss of generality. Assume that $\eta$ is a quantum operator molecule formula and has a probability operator term $\int\alpha$. Using the $Prob$ axiom $\vdash\int\alpha= \mathop{\sum}\limits_{A\in\llbracket\alpha\rrbracket}T_A$, we replace the probability operator term $\int\alpha$ by the projective operator terms $T_A$, $A\in\llbracket\alpha\rrbracket$. Then, we show that the quantum operator formula $\eta'$ after substitution is equivalent to $\eta$.\qed

Assume that $\eta$ is a quantum operator molecule formula, given a mapping $\sigma:X\rightarrow S(H_{qB})$, if $t_1\leq t_2\in\eta^{+}$ then $\sigma(t_1)\leq\sigma(t_2)$, and if $t_1\leq t_2\in\eta^{-}$ then  $\sigma(t_1)\not\leq\sigma(t_2)$, thus we say that $\eta$ is $\leq$-consistent.

Assume that $[G]$ is a sub-system operator formula, if there is a partition $S$ such that $S$ can interpret sufficiently the formula $[G]$ which is contained in a quantum operator molecule, we say $\eta$ is a $s$-satisfiable, denoted by $S\models_s\eta$.

\begin{lemma}{\bf \cite{[18]}} If $\eta$ is consistent, then $\eta$ is $s$-satisfiable.
\end{lemma}

\begin{theorem} If a quantum operator formula $\eta$ is consistent, then there is a quantum operator structure $M=<S(H_{qB}),V,S,\rho,\mu>$ such that $M\models\eta$.
\end{theorem}

\noindent {\bf Proof.}
We recall that every quantum operator formula is consistent if and only if one of quantum operator molecule formulae is consistent. So it suffices to consider a consistent quantum operator molecule formula. According to Proposition 5.3, we assume that $\eta$ contains no probability operator term $\int\alpha$.

Given a quantum operator molecule formula $\eta$ free of probability term, we consider a quantum operator molecule formula $\eta_1=\eta\sqcap(\mathop{\sum}\limits_{A\subseteq qB}T_A= I)$, and obtain a result that $\eta$ is consistent if and only if $\eta_1$ is consistent.

For any $G\in Alg(S)$, we give a quantum operator molecule formula  $\eta_1\sqcap(\sqcap_{G\in Alg(S)}[G])$, then obtain $\vdash\eta_1\equiv (\eta_1\sqcap(\sqcap_{G\in Alg(S)}[G]))$. According to the Axiom $Unit$, we have $\vdash([G])\sqsupset(\mathop{\sum}\limits_{A\subseteq G}T_A^G=I)$, so $\vdash\mathop{\bigsqcap}\limits_{G\in Alg(S)}([G])\sqsupset\mathop{\bigsqcap}\limits_{G\in Alg(S)}(\mathop{\sum}\limits_{A\subseteq G}T_A^G=I)$. Therefore, $\vdash\eta_1\sqsupset(\eta_1\sqcap\mathop{\bigsqcap}\limits_{G\in Alg(S)}(\mathop{\sum}\limits_{A\subseteq G}T_A=I_G))$ and so $\vdash\eta_1\equiv(\eta\sqcap\mathop{\bigsqcap}\limits_{G\in Alg(S)}(\mathop{\sum}\limits_{A\subseteq G}T_A=I_G))$. Let $\eta_2=\eta\sqcap(\mathop{\bigsqcap}\limits_{G\in Alg(S)}(\mathop{\sum}\limits_{A\subseteq G}T_A=I_G))$, thus
$\vdash\eta_2\equiv\eta_1$.

For any $G_1,G_2,A_1,A_2$ such that $G_1,G_2\in Alg(S)$, $A_1\subseteq G_1$, $A_2\subseteq G_2$, we consider a quantum operator molecule formula $\eta_2\sqcap\bigsqcap_{\substack{G_1,G_2\in Alg(s),A_1\subseteq G_1,A_2\subseteq G_2}}(T^{G_1\cup G_2}_{A_1\cup A_2}=T^{G_1}_{A_1}\otimes T^{G_2}_{A_2})$. According to the Axiom $MO1$, we obtain: $$\vdash\eta_2\equiv\eta_2\sqcap\bigsqcap_{\substack{G_1,G_2\in Alg(s),A_1\subseteq G_1,A_2\subseteq G_2}}(T^{G_1\cup G_2}_{A_1\cup A_2}=T^{G_1}_{A_1}\otimes T^{G_2}_{A_2}).$$ Let $\eta_3=\eta_2\sqcap\bigsqcap_{\substack{G_1,G_2\in Alg(s),A_1\subseteq G_1,A_2\subseteq G_2}}(T^{G_1\cup G_2}_{A_1\cup A_2}=T^{G_1}_{A_1}\otimes T^{G_2}_{A_2})$, thus $\vdash\eta_3\equiv\eta_2$. The same proves that $\vdash\eta^*\equiv\eta_3$, where $\eta^*=\eta_3\sqcap(T_\emptyset^\emptyset\equiv I)$.

In conclusion, we have $\vdash \eta\equiv\eta^* $. So, we have a result that $\eta$ is consistent if and only if $\eta^*$ is consistent.

In the following, we concern with the existence of quantum operator structure about $\eta^*$ which is consistent. $\eta^*$ is a quantum operator molecule formula which consists of equations about the projective operators, inequations about $t_1\leq t_2$ and sub-systems $[G]$. As a part of $\eta^*$, we suppose that $\eta^*_{\leq}$ is a conjunction form which consists of equations about the projective operators and inequations about $t_1\leq t_2$ in $\eta^*$. The remainder is denoted by $\eta^*_{[G]}$. We note that if $\eta^*_{\leq}$ is inconsistent, then $\eta^*_{\leq}\sqcup\eta^*_{[G]}$ is also inconsistent, hence $\eta^*$ is inconsistent. We denote $\eta^*_{\leq}(T^G_A/x_{T^G_A})$ by $\eta_0$ which is obtained from $\eta^*_{\leq}$ by replacing each term of the form $T^G_A$ by $x_{T^G_A}$, that is, $\eta_0=\eta^*_{\leq}\{T^G_A/x_{T^G_A}\}$. Similarly, we replace each variable terms $x_{T^G_A}$ by $T^G_A$ in $\eta_0$, and get $\eta^*_{\leq}=\eta_0\{x_{T^G_A}/T^G_A\}$. The $\eta_0$ is a quantum operator analytical formula of \textbf{sub-EQOL}. The following will show there is a quantum operator structure $M$ such that $M\models\eta_0$.

Proof by contradiction. if there is not any quantum operator structure $M$ such that $M\models\eta_0$, then $\boxminus\eta_0$ is a quantum operator analytical tautology, that is, $\vdash\boxminus\eta_0$ using $QTaut$. Furthermore, using $RCF$, then we have $\vdash\boxminus\eta_0\{x_{T^G_A}/T^G_A\}$, that is, $\vdash\boxminus\eta^*_{\leq}$. Hence, $\eta^*_{\leq}$ is inconsistent. Therefore, $\eta^*$ is inconsistent. Since $\eta\equiv\eta^*$, we also obtain that $\eta$ is inconsistent. This falls into conflict with the prerequisite that $\eta$ is consistent. Hence, there must be a quantum operator structure $M'$ such that $M'\models\eta_0$. Then, according to the definition $\eta_0$ and Lemma 5.1, there must be a quantum operator structure $M$ such that $M\models\eta$(or $M\models\eta^*$).\qed

In the following, we build this quantum operator structure $M=<S(H_{qB}),V,S,\rho,\mu>$ such that $M\models\eta$.

(1) $\rho_{\emptyset}=I$.

(2) For any $G\in S$, $\rho=\mathop{\otimes}\limits_{G\in S}\rho^{G}$.

(3) Define an assignment function $\sigma:X\rightarrow S(H_{qB})$ such that
$$\sigma(x_{T^G_A})=\left\{
          \begin{array}{ll}
            T^G_A, & \hbox{if $x_{T^G_A}$ is a variable operator term in $\eta_O $.} \\
            0, & \hbox{otherwise,}
          \end{array}
        \right.
$$
We construct $\mu=\{T^{G}_A\}_{A\subseteq G}$. Then, we define a quantum density operator $\rho^{G}=\mathop{\sum}\limits_{A\subseteq G}p_A\rho_A$, where $p_A=tr(T^A_{G}(\rho^{G}))$ and $\rho_A=|(\wedge A)_G\rangle\langle(\wedge A)_G|$.

\begin{example}
We consider the following quantum operator molecule formula $\eta$:
$$[qB]\sqcap(x\leq T^{qB}_{qB})\sqcap(\frac{1}{2}I\leq T^{qB}_{\emptyset}).$$
We build a quantum operator structure:
$$M=<S(H_{qB}),V,S,\rho,\mu>,$$ where $qB=\{qb_1,qb_2\}$, $V=\{00,01,10,11\}$, $S=\{\emptyset, qB\}$. We have $Alg(S)=\{\emptyset,qB\}$, and define an assignment function
\begin{center}
$\sigma:X\rightarrow S(H_{qB})$ such that $\sigma(x)=T^{qB}_{qB}$.
\end{center}
Let
$\mu=\{T_{\emptyset}^{\emptyset},T_{\emptyset}^{qB}, T_{qB}^{qB}\}$, we give a density operators as follows
$$\rho=0.6|00\rangle\langle 00|+0.4|11\rangle\langle 11|.$$
We can show that $M\models\eta$.
\end{example}

\begin{theorem} Exogenous quantum operator logic is complete, that is, if $\models\gamma$, then $\vdash\gamma$.
\end{theorem}

\noindent {\bf Proof.}
Assume that $\nvdash\gamma$, we have $\nvdash\boxminus(\boxminus\gamma)$ followed by Atom $QTaut$ and $QMP$, thus $\boxminus\gamma$ is consistent. Then by the Theorem 5.1, there is a quantum operator structure $M$  such that $M\models\boxminus\gamma$. So $M\not\models\gamma$. This falls into conflict with the prerequisite $\models\gamma$. Hence, the theorem is valid.\qed

\section{Application examples}

In this section, in order to see the usefulness of our logic, we consider several examples including Bell states, the BB84 protocol and quantum loop programs. In particular, we propose a novel notion of quantum Markov chain which can be used to describe quantum loop programs.

\subsection{Reasoning about Bell states}

The Bell states are firstly studied by Einstein, Podolsky and Rosen. They are a concept in quantum information system, represent the most simple example of entanglement, and have applied for designing quantum communication. An independent sub-system is said to be in Bell state which be composed of a pair of qubits if they are maximally entangled. For example, the following state is a Bell state: $$|\psi\rangle=\frac{1}{\sqrt{2}}(|10\rangle+|01\rangle).$$

In\cite{[13]}, reasoning about Bell states has been discussed using \textbf{EQPL}. By applying the meta-theorem theorem, one can use \textbf{EQPL} to derive that a pair of qubits in a Bell state is necessarily entangled.

By using our logic, the entanglement about a pair of qubits in a Bell state can be also derived. Specific details go as follows.

Given a Bell state $|\psi\rangle=\frac{1}{\sqrt{2}}(|10\rangle+|01\rangle)$, its corresponding density operator is represented as $$\rho=|\psi\rangle\langle\psi|=\frac{1}{2}[|10\rangle\langle 10|+
|10\rangle\langle 01|+|01\rangle\langle 10|+|01\rangle\langle 01|].$$
Assume that there is a pair of qubits $qB=\{qb_1,qb_2\}$, we have $\rho_{qb_1}=\frac{1}{2}I$, $\rho_{qb_2}=\frac{1}{2}I$. We denote $tr(P_v\rho)$ by $\rho(v)$, $v\in\{00, 01, 10,11\}$, and have$\rho(00)=\rho(11)=0$, $\rho(01)=tr(P_{01}\rho)=\frac{1}{2}$ and $\rho(10)=tr(P_{10}\rho)=\frac{1}{2}$. The following projective operators are necessary: $$T_{\emptyset}^{qB}=0, T_{qB}^{qB}=0, 0<T_{\{qb_1\}}^{qB}, 0<T_{\{qb_2\}}^{qB}, 0<T_{\{qb_1\}}^{\{qb_1\}}, 0<T_{\{qb_2\}}^{\{qb_2\}}.$$

The fact that a pair of qubits in the Bell state is entangled can be expressed as the following quantum operator formula of \textbf{EQOL}, denoted by $\eta$:
$$([\{qb_0,qb_1\}]\sqcap\gamma)\sqsupset(\boxminus[\{qb_0\}]\sqcap\boxminus[\{qb_1\}]).$$
where $\gamma:=(\gamma_1\sqcap\gamma_2\sqcap\gamma_3\sqcap\gamma_4)$, $\gamma_1:=(T_{\emptyset}^{qB}=0)$, $\gamma_2:=(T_{qB}^{qB}=0)$,
$\gamma_3:=(0<T_{\{qb_1\}}^{qB}=\frac{1}{2}I)$,
$\gamma_4:=(0<T_{\{qb_2\}}^{qB}=\frac{1}{2}I)$.

In the above formula, $[\{qb_0,qb_1\}]$ implies that the quantum system is an independent two qubit sub-system. $\gamma$ is a quantum operator sub-formula which is used to describe the Bell state $|\psi\rangle=\frac{1}{\sqrt{2}}(|10\rangle+|01\rangle)$. $\boxminus[\{qb_0\}]\sqcap\boxminus[\{qb_1\}]$ implies that neither $qb_0$ nor $qb_1$ forms an independent sub-system. Then, if we derive that $\eta$ is valid, we will be able to interpret that a pair of qubits in a Bell state is entangled. In other words, we will derive the following assertion:
$$\vdash([\{qb_0,qb_1\}]\sqcap\gamma)\sqsupset(\boxminus[\{qb_0\}]\sqcap\boxminus[\{qb_1\}]).$$

Reasoning about this assertion is as follows.
\\
\noindent {\bf Proof.}

(1) $[\{qb_1,qb_2\}]$. $\quad$P(recondition)

(2) $[\{qb_1\}]$. $\quad$H(ypothesis)

(3) $([\{qb_1,qb_2\}]\sqsupset([\{qb_1\}]\sqsupset[\{qb_2\}]))$. $\quad$ Axiom $Sub\cup$

(4) $([\{qb_1\}]\sqsupset[\{qb_2\}])$. $\quad$(1),(3)

(5) $[\{qb_2\}]$. $\quad$(2),(4)

(6) $\gamma=\gamma_1\sqcap\gamma_2\sqcap\gamma_3\sqcap\gamma_4$. $\quad$P

(7) $\gamma_2=T_{qB}^{qB}=0$. $\quad$(6)

(8) $([\{qb_1\}]\sqcap[\{qb_2\}])\sqsupset(T_{\{qb_1,qb_2\}}^{\{qb_1,qb_2\}}\equiv T^{\{qb_1\}}_{\{qb_1\}}\otimes T^{\{qb_2\}}_{\{qb_2\}})$. $\quad$ Axiom $MO1$

(9) $T^{\{qb_1\}}_{\{qb_1\}}\otimes T^{\{qb_2\}}_{\{qb_2\}}=0$. $\quad$(2),(5),(7)

(10) $(0<T^{\{qb_1\}}_{\{qb_1\}})\sqcap(0<T^{\{qb_2\}}_{\{qb_2\}})$. $\quad$ P

(11) $0<(T^{\{qb_1\}}_{\{qb_1\}}\otimes T^{\{qb_2\}}_{\{qb_2\}})$. $\quad$(10)

(12) $\bot$. $\quad$(9),(10)

(13) $\Bot$. $\quad$ Soundness.\qed

Therefore, by deduction, we obtain that this assertion $\eta$ is valid, and show that neither $qb_0$ nor $qb_1$ forms an independent sub-system. That is, this quantum system is entangled.

\subsection{Reasoning about BB84 protocol}
Quantum communication and cryptographic protocols are becoming an important practical technology. In a great number of research, their correctness has been proved using the methods of quantum computation and quantum information. But, few of them make use of formal methods such as formal model languages and logic deduction. In this subsection, we will reason about BB84 protocol using \textbf{EQOL}.

The BB84 developed by Bennett and Brassard in 1984\cite{[26]}, is a quantum cryptographic protocol based on the law of quantum mechanics. The basic BB84 protocol is as follows.

Assume that there are two groups of polarization basis(rectilinear basis and diagonal basis), and four polarizations(vertical, horizontal, diagonal and anti-diagonal).

(1) Alice chooses a random string of bits $\overline{A}$(polarization basis), and prepares a string of qubits $\overline{Q}$ with a random string of bits $\overline{K}_A$(polarization) such that belongs to the chosen basis.

(2) Alice sends this strings of qubits to Bob. For each qubit, Bob randomly chooses a polarization basis $\overline{B}$ and measures the polarization of qubit. Let $\overline{K}_B$ be the measure results.

(3) Alice and Bob use the public channel to compare their polarization bases, and determinate at which positions the polarization bases are equal, and keep only the polarization data at those positions. If no interferes of communication channel or eavesdroppings, these data should be the same. We call them raw keys.

(4) At the last step, Alice and Bob use some classical methods to check whether those raw keys are the same, otherwise, there exist errors and eavesdropping.

There is a method verifies that there are errors and eavesdropping in BB84 protocols. That is, at those positions that the polarization bases are equal, Bob would choose and announce a random subset of their keys, then Alice would compare this string of bits with one of her own at corresponding positions. Under the non-noise condition, if there is different between two strings of bits, then one would assert there is eavesdropping in this BB84 protocol, otherwise this protocol is efficient. Under the noise condition, if the bit error rate of two strings of bits reaches a certain threshold, then one would assert there is eavesdropping in this BB84 protocol. So this negotiation fails.

Generally, one can use algorithm to verify the BB84 protocol in quantum computation. In the following, we will use our logic to derive whether there is eavesdropping in the BB84 protocol after generating the raw keys.

Firstly, we build a quantum operator formula to represent an assert whether there is eavesdropping in the BB84 protocol.

Assume that $H_{qB}$ is a composite quantum system which is a tensor product of the $n$-dimensional Hilbert space $H_A$, $H_{K_A}$, $H_{B}$ and $H_{K_B}$. $qB_{\Delta}$ is a finite set of qubit symbols $\{qb_{\Delta}^i|i=1,2,\cdots,N\}$, $\Delta\in\{A,K_A,B,K_B\}$. For any $M\subseteq\{1,2,\cdots,N\}$, let $qB_{\Delta}(M)=\{qb_{\Delta}^i|i\in M\}$, we build an exogenous quantum operator formula as follows:
$$\varphi:=(0<\int(\wedge_{i\in M}(qb_A^i\leftrightarrow qb_B^i)))\sqsupset (aI\leq\int(\vee_{j\in M}(qb_{K_A}^j\leftrightarrow\neg qb_{K_B}^j))),$$
where $a>0$ is an arbitrary small number.

A detailed analysis of this formula go as follows. In this formula, one of classical formulas $$\wedge_{i\in M}(qb_A^i\leftrightarrow qb_B^i)$$
denotes that two groups of the polarization bases are equal. Then, $$0<\int(\wedge_{i\in M}(qb_A^i\leftrightarrow qb_B^i))$$
is a quantum operator atomic proposition. Given a density operator, if it is true, then it denotes that the probability of two groups of the polarization bases be equal is greater than zero. And, the other of classical formulas
$$\vee_{j\in M}(qb_{K_A}^j\leftrightarrow\neg qb_{K_B}^j)$$
denotes that two groups of raw keys have at least one pair of different raw keys. Then, the quantum operator axiom proposition
$$aI\leq\int(\vee_{j\in M}(qb_{K_A}^j\leftrightarrow\neg qb_{K_B}^j))$$
denotes that the probability about different raw key be exist is greater that the threshold $a$. That is, it is the extent of the bit error.

Then, the following quantum operator formula
$$\varphi:=(0<\int(\wedge_{i\in M}(qb_A^i\leftrightarrow qb_B^i)))\sqsupset (aI\leq\int(\vee_{j\in M}(qb_{K_A}^j\leftrightarrow\neg qb_{K_B}^j)))$$
denotes if it is possible that two groups of the polarization bases is equal, then the probability of the corresponding two groups of the raw keys be different is greater that the threshold $a$.

Hence, if we need to reason about there is eavesdropping in the BB84 protocol, then we will only derive that a given density operator $\rho$ satisfies $\varphi$, that is,
$$\rho\models\varphi.$$

The above analysis gives a method to reason about that there is eavesdropping in the BB84 protocol. Besides, in open environment, the communication process of BB84 protocol can be considered as a quantum Markov chain\cite{[25]}. In the future work, we will discuss the related properties about the BB84 protocol over quantum Markov chain such as several satisfiability problems.

\subsection{Quantum Markov chain based on \textbf{EQPL}}

Quantum Markov chain(\textbf{QMC}) is a mathematical formalism for the discrete-time evolution of open quantum systems. There are several versions of quantum Markov chains. In\cite{[27]}, a quantum Markov chain is a 2-tuple $<G,\varepsilon>$ where $G$ is a directed graph and $\varepsilon=[\varepsilon_{ij}]$ is a transition operator matrix. Every element $\varepsilon_{ij}$ labels the edge of from vertex $j$ to vertex $i$ where the sum of every column forms a quantum operator. In\cite{[28]}, a quantum Markov chain $<H,\varepsilon>$ is extended from a classical Markov chain of $<S,P>$ where the state space is replaced by a Hilbert space and its transition matrix is replaced by a super-operator. A similar type of \textbf{QMC} having the same power is given by $<S,Q,AP,L>$\cite{[25]}. In this model, $S$ is a finite set of classical states, $Q:S\times S\rightarrow SI(H)$ such that $\sum_{t\in S}Q(s,t)\eqsim I_{H}$ is called by a super-operator weighted Markov chain for each $s\in S$. The classical properties of states are described using classical label function $L:S\rightarrow 2^{AP}$ where $AP$ is a finite set of classical atomic propositions. However, the quantum properties can be not described in this model. In order to describe quantum properties, we introduce a novel notion of quantum Markov chain.

\begin{definition}
An exogenous quantum Markov chain is a five-tuple $$M_Q=<H_{qB},\varepsilon,l_{init},AP,L>.$$
where \emph{(1)} $\varepsilon$ is a quantum operator over $H_{qB}$, \emph{(2)} $l_{init}\subseteq H_{qB}$ is a Hilbert subspace of quantum initial states, \emph{(3)} $L$ is a labeling mapping from $D(H_{qB})$ to $2^{AP}$, where $AP\subseteq qAtom$, $|AP|=n$.
\end{definition}

The behaviour of quantum Markov chain can be described as follows: from $\rho_0\in l_{init}$, if the current state is in a density operator $\rho$, then it will be in the state $\varepsilon(\rho)$ in the following step. Meanwhile, the density opertaor $\rho_i$ after the $ith$ transtion is labeled by $L(\rho_i)\in 2^{AP}$. It is similar with the labeling function of classical Markov chain $<S,P,l_{init},AP,L>$\cite{[29]}. In the classical Markov chain, the labeling function value is a valuation of a classical formula which describes the properties of the current state. Similarily, in the exogenous quantum Markov chain, the labeling function value is a valuation of a quantum operator formula which describes the properties of the current state. Please note that the $AP$ of exogenous quantum Markov chain is a subset of quantum operator atomic propositions, but that of classical Markov chain is a set of classical atomic propositions.

According to the above analysis, our logic can describe the quantum properties of states over quantum Markov chains.

Given any $\rho\in D(H_{qB})$, the support $supp(\rho)$ of $\rho$ denotes a space spanned by eigenvectors of $\rho$ with non-zero eigenvalues. Let $\rho, \rho'\in D(H_{qB})$, we say that $\rho$ is adjacent to $\rho'$, written $\rho\rightarrow\rho'$, if $supp(\rho')\subseteq\varepsilon(supp(\rho))$. An infinite sequence $\pi=\rho_0\rightarrow\rho_1\rightarrow\cdots$ of adjacent density operators is callled a path from initial states $\rho_0\in l_{init}$. We denote all infinite paths from $\rho_0$ as $Paths(\rho_0)$. The following state or paths is useful.

(1) The $i-th$ quantum state of path $\pi$: $\pi[i]$.

(2) $\pi[i..]$ for $\rho_i\rho_{i+1}\cdots$.

(3) $\pi[..i]$ for $\rho_0\rho_{1}\cdots\rho_i$.

Reachability analysis is an important issue in model checking\cite{[19]}. We will focus on four reachability properties: future reachability, global reachability, infinitely often reachability, ultimately forever reachability.

(1) future reachability: $\pi\models F\gamma$ if and only if $\exists i\geq0$, $\pi[i]\models\gamma$,

(2) global reachability: $\pi\models G\gamma$ if and only if $\forall i\geq0$, $\pi[i]\models\gamma$,

(3) infinitely often reachability: $\pi\models U\gamma$ if and only if $\exists i\geq0$, $\forall j\geq i$, $\pi[j]\models\gamma$,

(4) ultimately forever reachability: $\pi\models I\gamma$ if and only if $\forall i\geq0$, $\exists j\geq i$, $\pi[j]\models\gamma$.

\begin{definition}
Given an exogenous quantum Markov chain $M_Q,\rho_0\in l_{init}$, $\Delta=\{F,G,U,I\}$, $\gamma$ is an exogenous quantum operator formula, then we define $M_Q,\rho_0\models\Delta\gamma$ if and only if for any $\pi\in Paths(\rho_0)$, $\pi\models\Delta\gamma$.
\end{definition}

\subsection{Reachability of generalized quantum loop programs}

Recently, quantum loop programs have attracts a few author's attention\cite{[30],[31],[32],[33],[34]}. The paper\cite{[31]} has given several criteria for deciding termination of a quantum loop on a given input. Meanwhile, quantum loop program in the open environment has been proposed, called by generalized quantum loop program(\textbf{GQLoop}). The results show that \textbf{GQLoop} can be modeled by quantum Markov chain\cite{[35]}. In the following, in order to reason about the termination of \textbf{GQLoop}, we will model \textbf{GQLoop} by using exogenous quantum markov chain, describe the termination property as a quantum operator formula.

Suppose that we have a quantum system which has $n$ quantum registers $qb_1,qb_2,\cdots,qb_n$, and each of their state spaces is $H_i$, $i\leq n$. We define a quantum operator $K:D(H_{qB})\rightarrow D(H_{qB})$ on a tensor product space $H_{qB}=\otimes_{i=1}^nH_i$, i.e., $K(\rho)=\Sigma_{i=1}^dE_i\rho E_i^{\dagger}$, where $\{E_i\}$ is a collection of operation elements satisfied $\Sigma_{i=1}^dE_i^{\dagger}E_i=I$, $\rho\in D(H_{qB})$, $d=dim(H_{qB})$. Let $M=\Sigma_mmM_m$ be an observable quantity over $H_{qB}$, we write $spec(M)=\{m\}$ for the spectrum of $M$. For any $X\subseteq spec(M)$, we introduce a kind of generalized quantum loop program\cite{[31],[32],[33],[34],[35]} defined by $K$,$M$ and $X$ may be written as follows:
\begin{equation}
  while(M[\overline{q}]\in X)\{\overline{q}:=K(\overline{q})\},
\end{equation}
where $\overline{q}$ is a sequence $qb_1,qb_2,\cdots,qb_n$ of quantum registers. Assume that $M_1=M_X=\Sigma_{m\in X}M_m$, $M_0=M_{\overline{X}}=I-M_X=\Sigma_{m\in spec(M)}$ $_{-X}M_m$ and $I$ a unit operator over $H_{qB}$, the guard $``M\in X"$ in formula(1) implies that the projective measurement $M_X$,$M_{\overline{X}}$ is applied to $\overline{q}$. The work and computational process of the \textbf{GQLoop} can be visualized by \textbf{Figure 1} and \textbf{Figure 2}.
\begin{center}
\includegraphics[width=0.82\textwidth]{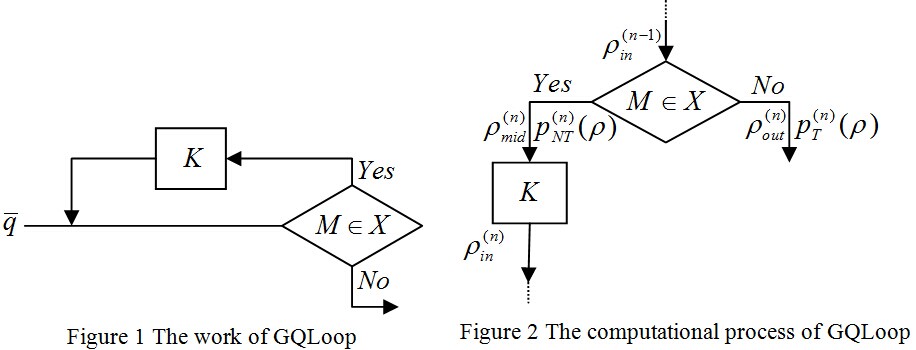}
\end{center}
For any input state $\rho^{(0)}_{in}=\rho_0\in D(H_{qB})$, if \textbf{GQLoop} doesn't terminate before the $n-1$ steps, then \textbf{GQLoop} will terminate with probability $p_T(\rho^{(n-1)}_{in})=tr(M_{\overline{X}}\rho^{(n-1)}_{in})$ and the output state $\rho^{(n)}_{out}=\frac{M_{\overline{X}}\rho^{(n-1)}_{in}M_{\overline{X}}}{p_T(\rho^{(n-1)}_{in})}$ in the $n-th$ step. In the case of nontermination, \textbf{GQLoop} will continue with probability $p_{NT}(\rho^{(n-1)}_{in})=1-p_{T}(\rho^{(n-1)}_{in})=tr(M_X\rho^{(n-1)}_{in})$, the nontermination state $\rho^{(n)}_{mid}=\frac{M_X\rho^{(n-1)}_{in}M_X}{p_{NT}(\rho^{(n-1)}_{in})}$, and the next input state $\rho^{(n)}_{in}=K(\rho^{(n)}_{mid})$. For convenience, we denote the termination probability and the nontermination probability in the $n-th$ step by $p^{(n)}_{NT}(\rho)$, $p^{(n)}_{T}(\rho)$, respectively.

Given an input state $\rho^{(0)}_{in}\in D(H_{qB})$ and a generalized quantum loop program, for any $n\in N$, if $p^{(n)}_{NT}(\rho^{(0)}_{in})=0$, we say that the generalized quantum loop program terminates on input state $\rho^{(0)}_{in}$.

A generalized quantum loop program is an exogneous quantum Markov chain. In fact, a generalized quantum loop program has been expressed as a quantum Markov chain\cite{[34],[35]}. We give a set of quantum operator atomic propositions $AP=\{pI\leq T^{qB}_{X},T^{qB}_{X}\leq pI|p\in[0,1]\}$, where the $pI\leq T^{qB}_{X}$($T^{qB}_{X}\leq pI$) implies that the probability is greater(less) or equal to $p$ that the measurement outcome is in $X$. Then, the guard $M\in X$ can be written as a quantum operator formula $\gamma\equiv(pI\leq T^{qB}_{X})\sqcap(T^{qB}_{X}\leq pI)$(or $\gamma\equiv(T^{qB}_{X}=pI)$), $p\in [0,1]$. Given an input state $\rho^{(0)}_{in}$, after $n$ steps, if $\rho^{(n)}_{in}\models\gamma$ or $\rho^{(n)}_{in}\models(T^{qB}_{X}=pI)$, then the formula $\gamma$ denotes that the probability of the measurement outcome being in $X$ is $p$ in the state $\rho^{(n)}_{in}$, or the \textbf{GQLoop} doesn't terminate with a probability of $p$. Besides, using the definition of \textbf{GQLoop}, we get $p=p^{(n)}_{NT}(\rho^{(0)}_{in})=tr(T^{qB}_{X}(\rho^{(n)}_{in}))$.

\begin{proposition}
Given any input state $\rho^{(0)}_{in}$, there is a positive integer $n$ such that a generalized quantum loop program terminates after $n$ steps, that is, $p^{(n)}_{NT}(\rho^{(0)}_{in})=0$, if and only if there is an exogenous quantum Markov chain $M_Q$ such that $M_Q,\rho^{(0)}_{in}\models F\gamma$, where $\gamma\equiv(T^{qB}_{X}=O)$.
\end{proposition}

In the above proposition, $M_Q,\rho^{(0)}_{in}\models F\gamma$ implies that for any $\pi\in Paths(\rho^{(0)}_{in})$, if $\pi\models F\gamma$, then there is a positive integer $n$ such that $\pi[n]=\rho^{(n)}_{in}\models\gamma$. Therefore, we have $tr(T^{qB}_{X}(\rho^{(n)}_{in}))=0$ or $p^{(n)}_{NT}(\rho^{(0)}_{in})=0$, that is, the probability of termination is 1.

This proposition shows that the termination of \textbf{GQLoop} can be turned into solving the satisfiability problem of the formula $F\gamma$.

\section{Conclusion}

The main contribution of this paper is to introduce a novel logic for open quantum systems by using exogenous approach where the density operator is considered from the state logic point of view. We call it an exogenous quantum operator logic(\textbf{EQOL}). It is an extension of exogenous quantum propositional logic(\textbf{EQPL}). The main difference between \textbf{EQPL} and \textbf{GQLoop} is that the former use the unit vectors to describe the closed quantum systems, whereas the latter use the density operators to describe the open quantum systems.

The main idea is to replace the term languages of \textbf{EQPL} by the operator languages in our logic, and interpret them in the super-operators. Through using the classical formulae and the operator terms, we recursively build quantum operator formulae. This approach is expressive enough to reason about open quantum systems where the density operators are considered.

As an axiomatic logical system, we present several axioms and rules. We show it is sound and complete. To illustrate the expressiveness of our logic, we cite some examples, for example, reasoning about the entanglement of the Bell states. The properties described in these examples are modeled by quantum operator formulae and reasoned about their satisfiability. Besides, we also use our logic to propose a novel notion of quantum Markov chain: exogenous quantum Markov chain, that be used to formalize the discrete-time evolution of open quantum systems. As its application, we illustrate that a generalized quantum loop programs can be described by an exogenous quantum Markov chains, its termination problems can be modeled and checked.

Along exogenous quantum operator logic, we still have much work to be done. As one of the future directions we are pursuing, we would be interesting to have a temporal version of exogenous quantum operator logic, for example \textbf{LTL} and \textbf{CTL}. Meanwhile, we also consider their SAT and model-checking problems. In order to describe the evolution of open quantum systems, we also plan to research the dynamic version of \textbf{EQOL}.

\section{Acknowledgements}

This work was partially supported by National Science Foundation of China(Grant Nos:11271237, 61228305)and the Higher School Doctoral Subject Foundation of Ministry of Education of China(Grant No:20130202110001).

\end{document}